\documentstyle[11pt,rpmaaspp4,epsf,rpmflushrt]{article}

\def\lta{\lesssim}
\def\gta{\gtrsim}

\def\kms{\>{\rm km}\,{\rm s}^{-1}}
\def\pc{\>{\rm pc}}

\def\Msun{\>{\rm M_{\odot}}}
\def\yr{\>{\rm yr}}

\def\Mdark{M_{\bullet}}
\def\rh{r_{\rm h}}

\def\r#1{{\tiny\raise5pt\hbox{\cite{#1}}}\hskip0pt}
\def\rr#1{{\tiny\raise5pt\hbox{,\cite{#1}}}\hskip0pt}
\def\rbar{{\tiny\raise5pt\hbox{--}\hskip0pt}}

\begin{document}

\thispagestyle{empty}

%%%%%%%%%%%%%%
% Initial Comment
%%%%%%%%%%%%%%

\null
\vspace{-1.8truecm}
\hrule
\vspace{-0.16truecm}
\noindent NATURE 385, 610--612.\hfill LETTER TO NATURE\par
\hrule
\bigskip\smallskip

%%%%%%%%%%%%%%%
% Title and authors
%%%%%%%%%%%%%%%

\baselineskip=20pt

{\Large\bf\noindent A massive black hole at the centre of the quiescent 
galaxy M32}

\baselineskip=15pt

{\large\bf\noindent 
Roeland P.~van der Marel\altaffilmark{1},
P.~Tim de Zeeuw\altaffilmark{2},
Hans--Walter Rix\altaffilmark{3} \&\\
Gerald~D.~Quinlan\altaffilmark{4}} 

\baselineskip=10pt

{\small\noindent $^1$Institute for Advanced Study, Olden Lane, Princeton,
New Jersey 08540, USA}\newline
{\small\noindent $^2$Sterrewacht Leiden, Postbus 9513, 2300 RA Leiden, The
Netherlands}\newline
{\small\noindent $^3$Steward Observatory, University of Arizona, Tucson, 
Arizona 85721, USA}\newline 
{\small\noindent $^4$Dept.~of Physics and Astronomy, Rutgers University, 
PO Box 849, Piscataway, New Jersey 08855, USA}

\bigskip

{\small\sl\noindent received 23 July 1996; accepted 10 January 1997;
published 13 February 1997.}

\bigskip

\baselineskip=13pt

%%%%%%%%%%%%%%%
% Abstract
%%%%%%%%%%%%%%%

\looseness=-2
\noindent {\bf Massive black holes are thought to reside at the
centres of many \hbox{galaxies\r{Ree90}\rr{Hae93},} where they power
quasars and active galactic nuclei. But most galaxies are quiescent,
indicating that any central massive black hole present will be starved
of fuel and therefore detectable only through its gravitational
influence on the motions of the surrounding stars. M32 is a nearby,
quiescent elliptical galaxy in which the presence of a black hole has
been \hbox{suspected\r{Kor95}\rbar\r{Ben96};} however, the limited
resolution of the observational data and the restricted classes of
models used to interpret this data have made it difficult to rule out
alternative explanations, such as models with an anisotropic stellar
velocity distribution and no dark mass or models with a central
concentration of dark objects (for example, stellar remnants or brown
dwarfs). Here we present space-based high-resolution optical spectra
of M32, which show that the stellar velocities near the centre of this
galaxy exceed those inferred from previous ground-based
observations. We use a range of general dynamical models to determine
a central dark mass concentration of $(3.4 \pm 1.6) \times 10^6$ solar
masses, contained within a region only $0.3\pc$ across. This leaves a
massive black hole as the most plausible explanation of the data,
thereby strengthening the view that such black holes exist even in
quiescent galaxies.}

\bigskip

%%%%%%%%%%%%%%%
% Main text
%%%%%%%%%%%%%%%

We observed M32 with the Faint Object Spectrograph (FOS) on the Hubble
Space Telescope (HST), on August 22, 1995. The COSTAR optics corrected
the spherical aberration of the HST primary mirror. Spectra of the
wavelength range 4572--6821~{\AA} were taken through square apertures
of $0.086''$ and $0.215''$, aligned on the nucleus and along the
photometric major axis. The kinematical analysis was done by fitting
each galaxy spectrum with the convolution of a template spectrum and a
gaussian line-of-sight velocity profile. The details of the
observations will be discussed by van der Marel, de Zeeuw \& Rix
(manuscript in preparation). The inferred kinematics are compared with
the best ground-based \hbox{data\r{Ben96}\rr{vdM94}} in Figure~1. The
rotation curve derived from the HST spectra rises more steeply than
that measured from the ground. The velocity dispersion from the HST
aperture closest to the centre is $156 \pm 10 \kms$.  The average of
the four independent dispersion measurements inside the central
$0.1''$ is $126 \kms$, exceeding significantly the 85--95$\kms$
measured from the ground.  \looseness=-2

Dynamical models are needed to infer the mass distribution from the
kinematical data, and to test whether the observed velocities are
consistent with a keplerian rise around a black hole. To this end we
developed a new technique to construct fully general axisymmetric
models that fit a given set of observations (Rix et al., van der Marel
et al., Cretton et al., manuscripts in preparation). It is based on
Schwarzschild's \hbox{approach\r{Sch79}}, and can be viewed as an
axisymmetric generalization of the spherical technique of Richstone
and \hbox{collaborators\r{Dre88}\rr{Ric90}\rr{Ric85}}. It consists of
four steps: (i) an axisymmetric mass density is chosen that fits the
M32 HST \hbox{photometry\r{Lau92}} after projection; (ii) the
gravitational potential is calculated, including a nuclear dark mass;
(iii) a representative library of orbits is calculated; and (iv) a
non-negative least-squares fit is performed to determine the
combination of orbits that reproduces the assumed density and best
fits the observational constraints.\looseness=-2

Previous studies of M32 have used either axisymmetric models with
distribution functions depending only on the two classical integrals
of motion (the energy $E$ and angular momentum $L_z$ around the
symmetry axis), $f(E,L_z)$, or spherical models with fully general
distribution functions. Models of the first type take into account the
flattening of M32 (projected axial ratio $\sim 0.73$), but have a
special dynamical structure. They successfully fit the ground-based
data when a nuclear dark mass is
\hbox{included\r{Qia95}\rbar\r{Ben96},} but they leave open the
possibility that more general models might fit the data without a dark
mass. Models of the second type allow for general anisotropy, but
ignore the flattening of M32. These spherical models were used only to
interpret some of the older M32 \hbox{data\r{Dre88},} and could not
fit these without a nuclear dark \hbox{mass\r{Ric90}.} However, we
found that our less restricted axisymmetric models can easily fit
these same data without a dark mass. So even though previous studies
have shown that existing data are consistent with a nuclear dark mass
in M32, it has not been shown that such a dark mass is
required.\looseness=-2

To improve this, we used our general axisymmetric technique to
determine which models can fit both the new HST data and the most
recent ground-based \hbox{data\r{Ben96}\rr{vdM94}}. The models take
the observational characteristics into account, and include velocity
profile shape data (not available from the HST because of the limited
spectral resolution of the FOS). Figure~2a shows a contour plot of the
$\chi^2$ of the fit for edge-on models with a dark central point mass.
The free parameters are the V-band mass-to-light ratio $\Upsilon_{\rm
V}$ of the stellar population, and the dark mass $\Mdark$. The models
labelled A--D are the best-fitting models for four fixed values of
$\Mdark$. Figure~3 compares the predictions of these models with the
observed major axis kinematics for each of the data sets. Models
without a dark mass are ruled out; even the best-fitting model without
a dark mass, Model~A, fails to fit the nuclear HST velocity
dispersions by $>40 \kms$. Model~C has $\Mdark = 3.4 \times 10^6
\Msun$, and provides the overall lowest $\chi^2$. However, we would
not necessarily expect our algorithm to constrain a single, best-fit
choice of the potential, since a three-dimensional distribution
function might have the freedom to adjust to changes of the potential
without affecting the fit to the data. Numerical experiments showed
that the structure in the $\chi^2$ plot near the two minima could in
fact be unreliable. We therefore do not claim that the local minimum
labelled C is necessarily special, but conservatively estimate the
allowed range for $\Mdark$ to be $(3.4 \pm 1.6) \times 10^6 \Msun$.
The uncertainty takes into account not only the observational errors
in the data, but also the possible influence of small ($<2 \kms$ in
the projected kinematics) numerical errors in the models (due to, for
example, gridding and discretization). M32 need not be edge-on, but
models constructed for inclination $i=55^{\circ}$ gave similar results
and the same allowed mass range. A full exploration of the range of
possible inclinations is computationally prohibitive, but our results
and previous \hbox{work\r{Deh95}} suggest that the inferred $\Mdark$
depends only weakly on the assumed inclination. Allowing for possible
triaxiality is not likely to change these conclusions; they are due
mostly to the radial behaviour of the potential, and not to its
shape. The best-fitting models are not $f(E,L_z)$ models, but are
similar in that they are tangentially anisotropic and require a
similar dark mass.  \looseness=-2

To constrain the size of the dark mass, models were built with the
point-mass potential replaced by a Plummer potential with the same
mass, but with scale length $\epsilon$. These were restricted to
$f(E,L_z)$ models for simplicity; constraints on $\epsilon$ from the
more general models should not differ by much. Figure~2b shows a
contour plot of the $\chi^2$ of the fit to the HST velocity
dispersions, for edge-on models. The best fit has $\epsilon = 0$ (a
point mass); models with $\epsilon \gta 0.06''$ are ruled out.

The mass concentration thus has a half-mass radius $\rh = 1.30
\epsilon \lta 0.08''$, implying a central density $\gta 10^8 \Msun
\pc^{-3}$ ($1'' = 3.39\pc$) and a mass-to-light ratio $\gta 20$ inside
$0.1''$. It cannot be a cluster of ordinary stars; that would evolve
rapidly by stellar \hbox{collisions\r{Lau92},} and a change in
mass-to-light ratio from $\sim 2$ at $1''$ to $\gta 20$ at $0.1''$
would yield large broad-band colour gradients, ruled out by
\hbox{observations\r{Lug92}\rr{Cra93}.} It must be dark, either a
massive black hole or a cluster of dark objects (e.g., stellar
remnants, brown dwarfs, planets).  No theory predicts the formation of
such clusters---except \hbox{perhaps\r{Mor93}\rr{Lee95}} for a cluster
of small black holes of mass $m \simeq 5$--$10\Msun$, each---and most
would not last for the age of the galaxy. A dark cluster is not an
acceptable alternative to a massive black hole if it will collapse to
a black hole in a short time, or expand or eject its mass through
explosions or evolve in some other way that would make it observable
(for then we would have to assume that it formed recently). Goodman
and \hbox{Lee\r{Goo89}} considered possible dark clusters for M32, and
argued that the clusters required $\rh \gta 0.1''$ to be acceptable;
these were consistent with ground-based data, which
\hbox{allowed\r{Ric90}} $\epsilon$ to be as large as $0.4''$. We
summarize and update the arguments here to show that all but the most
implausible clusters are now ruled out.

\looseness=-2 A cluster of stellar remnants of mass $m\gta 1\Msun$
would undergo core collapse in a short time, $5.5\times10^8\yr\,
\allowbreak (\!\Msun/m) \allowbreak (\rh/0.08'')^{3/2}$ for a Plummer
model (and shorter for more concentrated
\hbox{models\r{Qui96}}). Neutron stars or small black holes would form
binaries and merge during the collapse by dissipating energy through
gravitational \hbox{radiation\r{Qui89}.} The binaries would not be
able to stop the collapse until they had grown large through multiple
mergers, and even then they would at most cause rapid oscillations in
the central density, during which they would continue to
\hbox{grow\r{Lee95}.} One black hole would probably grow much faster
than the others in a runaway \hbox{manner\r{Lee93}.} Its mass would be
modest at first, perhaps $10^3 \Msun$, and might remain considerably
smaller than $3 \times 10^6 \Msun$. This model is not ruled out, but
still requires a black hole (although embedded in a dark cluster) to
fit the data. A cluster of white dwarfs would evolve similarly if the
white dwarfs collapse to neutron stars upon merging; if they explode
as supernovae instead the cluster would either lose mass or become
observable.

\looseness=-2 
A cluster of brown dwarfs or planets would have a long
collapse time, but a short collision \hbox{time\r{Goo89}}. The outcome
of a collision depends on the ratio of the collision and binding
\hbox{energies\r{Ben87}}: brown dwarfs of mass $m\gta0.05\Msun$ would
merge with little mass loss, leading again to the growth of a massive
object; planets or small brown dwarfs would disintegrate, leading to a
massive gas cloud that would be unlikely to remain dark.  The
constraint that the collision time be longer than the age of the
galaxy rules out clusters with $m\lta 0.03\Msun$.  It might
\hbox{allow\r{Goo89}} a cluster of brown dwarfs just below the
hydrogen-burning limit ($m\simeq0.08\Msun$) if $\rh\simeq0.08''$, but
another \hbox{argument\r{Qui96}} rules that out.  The luminous stars
of mass $m_*$ that pass through the cluster would get trapped by
dynamical friction in a time that depends only on the cluster density
and not on $m$ (if $m\ll m_*$), $5.5\times10^8\yr\, \allowbreak
(\!\Msun/m_*) \allowbreak (\rh/0.08'')^{3/2}$ (the collapse time given
above reduced by $m/m_*$).  The stars would sink to the centre,
probably to merge into a massive object because binaries would be
unable to eject stars from the deep potential well.  A cluster with
$\rh\simeq0.08''$ would trap nearly 0.1\% of the luminous stars, too
many to be acceptable. This same argument limits the size of a bizarre
dark cluster of elementary particles or tiny black holes to $\rh \lta
0.01''$.

The simplest and most plausible interpretation of the data is thus a
single massive black hole.  The X-ray emission from the centre of M32
could be from accretion onto the black hole, although its flux is low
enough to be from a low-mass X-ray \hbox{binary\r{Esk96}.} The absence
of a larger flux is \hbox{explicable\r{Kor95}:} stars will be tidally
\hbox{disrupted\r{Goo89}} only once every $10^3$--$10^4\yr$, and the
luminosity from continuous gas accretion can be
\hbox{suppressed\r{Nar95}.}

\looseness=-2
The arguments against a dark cluster in M32 rely mainly on the high
density it would have ($\gta 10^8$ ${\rm M}_{\odot} \pc^{-3}$).  These
arguments are weaker for the active galaxies M87 and NGC 4261, for
which the dark matter density required by HST gas
\hbox{kinematics\r{Har94}\rr{Fer96}} is only $10^5$ (similar to the
density of {\it luminous} material observed in some galactic nuclei
and globular clusters). Stronger arguments can be made for three other
black hole candidate galaxies requiring high densities: $10^{7.3}$ for
the quiescent galaxy NGC 3115, as derived from HST stellar
\hbox{kinematics\r{Kor96};} $10^{9.8}$ for our own Galaxy, as derived
from near-infrared velocity measurements of individual
\hbox{stars\r{Eck96};} and $10^{9.6\hbox{--}12.6}$ for the active
galaxy NGC 4258, as derived from water maser
\hbox{observations\r{Miy95}\rr{Mao95}.} Together with our M32 results
this provides compelling evidence for massive black holes in both
active and quiescent galaxies.

\bigskip\bigskip

%%%%%%%%%%%%%%%
% Acknowledgments
%%%%%%%%%%%%%%%

\small
\baselineskip=10pt

\noindent Acknowledgements: We thank Nicolas Cretton, Lars Hernquist,
Steinn Sigurdsson and Simon White for collaboration in various stages
of this project. The results are based on observations with the
NASA/ESA Hubble Space Telescope obtained at the Space Telescope
Science Institute. STScI provided financial support through a GO grant
and through a Hubble Fellowship awarded to RPvdM. STScI is operated
under NASA contract by the Association of Universities for Research in
Astronomy, Inc.

\bigskip

\noindent Correspondence should be addressed to R.P.v.d.M (email:
marel@sns.ias.edu).

\bigskip

%%%%%%%%%%%%%%%
% References
%%%%%%%%%%%%%%%

\clearpage

%%%%%%%%%%%%%%%
% Set the figure caption texts in boxes
%%%%%%%%%%%%%%%

\def\figcapone{Stellar rotation velocities $V$ (top panel) and
velocity dispersions $\sigma$ (bottom panel) in M32, as function of
major axis distance. The new HST data (aperture sizes indicated in the
figure) are compared with ground-based data obtained at the William
Herschel \hbox{Telescope\r{vdM94}} (WHT; resolution $\sim 0.8''$;
connected by a line for illustration; error bars smaller than plot
symbols) and the Canada-France Hawaii \hbox{Telescope\r{Ben96}} (CFHT;
resolution $\sim 0.5''$). The HST data were obtained with the G570H
grating on the FOS red detector, and were calibrated using arc lamp
spectra obtained during Earth occultations. The pointing accuracy was
limited by target acquisition uncertainties ($\sim 0.02''$) and
thermal effects on the spacecraft and Fine Guidance Sensors
(increasing to $\sim 0.1''$ during the 14 hours of the
observations). The actual aperture position for each observation was
calculated {\it post hoc} to $\sim 0.01''$ accuracy from models based
on HST \hbox{photometry\r{Lau92}} for the target acquisition data, the
observed count rate in each spectrum, and a verification image
obtained at the end of the observations. The template spectrum used
for the kinematical analysis was a mix of ground-based spectra of
stars of different spectral types, chosen to best match the M32
spectrum. The kinematical results were corrected for
line-spread-function differences between the galaxy and template
spectra. The line spread function for each galaxy spectrum was
calculated from models for the distribution of the galaxy light within
the aperture, taking into account the grating broadening function and
the finite size of the detector diodes. The relatively low velocity
dispersion inferred for the HST aperture second-closest to the nucleus
might have been affected by a systematic error, but none could be
identified.\looseness=-2}

\def\figcaptwo{{\sl a,} Contours of the $\chi^2$ that measures the
quality of the fit to the combined HST and ground-based kinematical
M32 data, for edge-on models with a fully general dynamical
structure. The model parameters $\Mdark$ and $\Upsilon_{\rm V}$ are
the dark nuclear point mass and the V-band stellar mass-to-light
ratio. The dots indicate models that were calculated. The predictions
of the models labelled A--D are shown in Fig.~3. The contours were
obtained through spline interpolation.  The first three contours
define the formal $68.3$\%, $95.4$\% and (heavy contours) $99.73$\%
confidence regions. Subsequent contours are characterized by a factor
two increase in $\Delta \chi^2 = \chi^2 - \chi_{\rm min}^2$. The
confidence values assume that there are only gaussian random errors in
the data, and no numerical errors in the models. In reality there are
numerical errors in the models, but they are known to be small (see
the legend of Figure~3). Tests show that these errors can be
responsible for the presence of the second (local) minimum in the
$\chi^2$ contours (although this might also be real), but that they
cannot make the predictions of models with $\Mdark < 1.8 \times 10^6
\Msun$ or $\Mdark > 5.0 \times 10^6 \Msun$ consistent with the data at
the formal $99.73$\% confidence level. We therefore conclude that
$\Mdark = (3.4 \pm 1.6) \times 10^6 \Msun$. {\sl b,} Contours of the
$\chi^2$ that measures the quality of the fit to the HST velocity
dispersions, for edge-on $f(E,L_z)$ models with an extended dark
nuclear object with scale length $\epsilon$ (and with $\Upsilon_{\rm
V}=2.51$ to best fit the data outside the central arcsec for this
assumed dynamical structure). A point mass ($\epsilon=0$) provides the
best fit; models with $\epsilon \gta 0.06''$ are ruled out.}

\def\figcapthree{Predicted rotation velocities $V$ and velocity
dispersions $\sigma$ of models~A--D in Figure~2, compared with the
major axis data from the HST, CFHT and WHT. Data points are plotted
equidistantly along the abscissa. Corresponding major axis distances
are given in the bottom panel. The WHT measurements are averages of
data obtained at positive and negative radii. The displayed quantities
form only a subset of all the data that were fitted (for the
ground-based \hbox{observations\r{Ben96}\rr{vdM94}}, velocity profile
shape data and position angles other than the major axis are also
available). Model~C is the edge-on model with the lowest $\chi^2$, and
has a dark nuclear point mass $\Mdark = 3.4 \times 10^6
\Msun$. Models~B and~D are (approximately) the best-fitting models for
$\Mdark=1.9$ and $5.4 \times 10^6 \Msun$, respectively. Model~A is the
best fit without a dark mass. This model is ruled out. It is already
marginally ruled out by the ground-based CFHT and WHT data, but only
the HST velocity dispersion measurements make the case clear-cut. The
accuracy of the model predictions was assessed by having them
reproduce the known results for $f(E,L_z)$
\hbox{models\r{Qia95}\rr{Deh95}.} From these tests we conclude that
numerical errors in our models are sufficiently small not to be of
influence to the conclusions of our paper (they are $<2 \kms$ in the
projected kinematics and $<0.01$ in the Gauss-Hermite
\hbox{moments\r{vdM94}} of the velocity profiles).}

%%%%%%%%%%%%%%%
% Figures
%%%%%%%%%%%%%%%

%%% FIGURE 1 %%%

\begin{figure}
\epsfxsize=8.0truecm
\centerline{\epsfbox{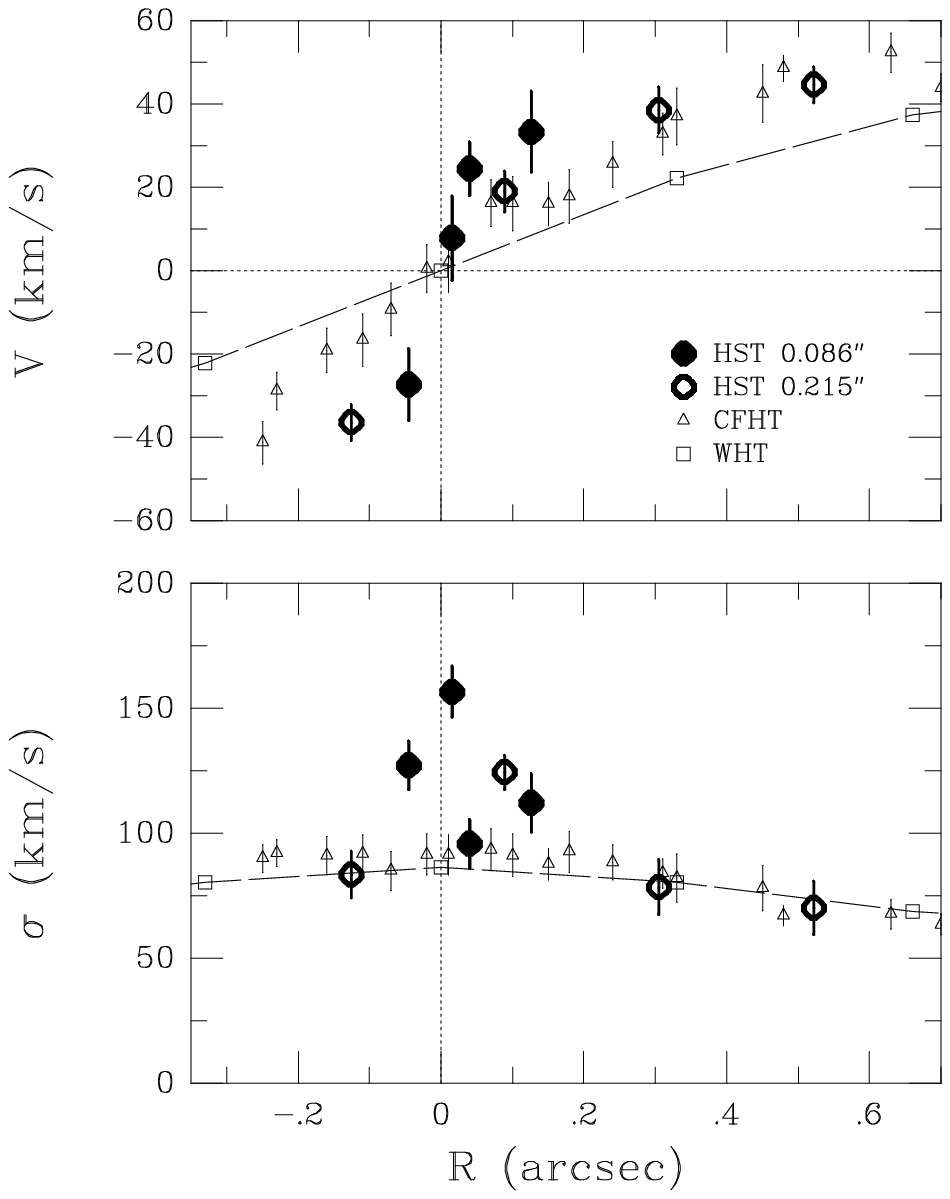}}
\caption{\small\figcapone}
\end{figure}

%%% FIGURE 2 %%%

\begin{figure} 
\noindent\begin{minipage}[b]{8.0truecm}
\epsfxsize=8.0truecm
\epsfbox{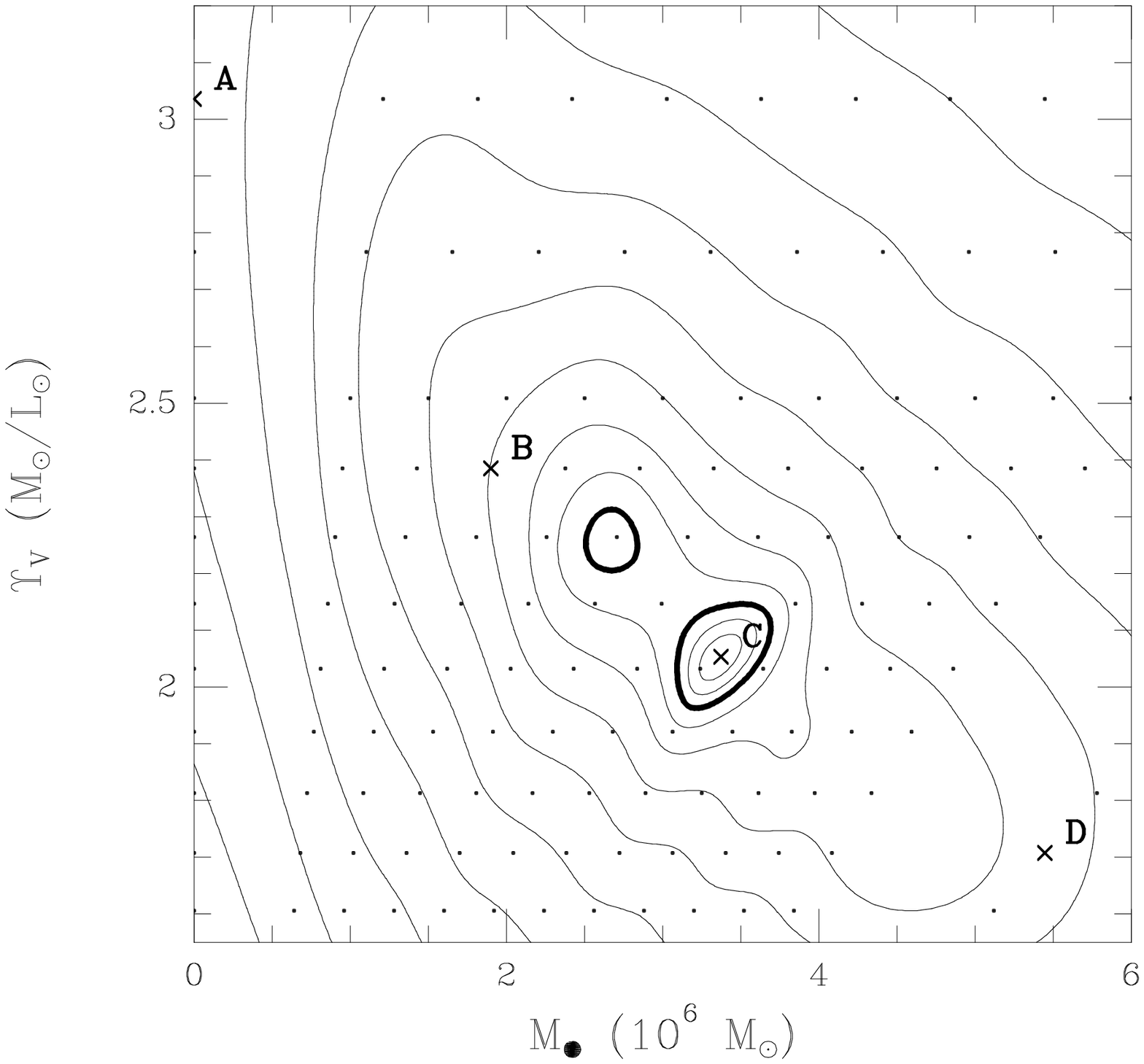}
\end{minipage}
\hfill
\noindent\begin{minipage}[b]{8.0truecm}
\epsfxsize=8.0truecm
\epsfbox{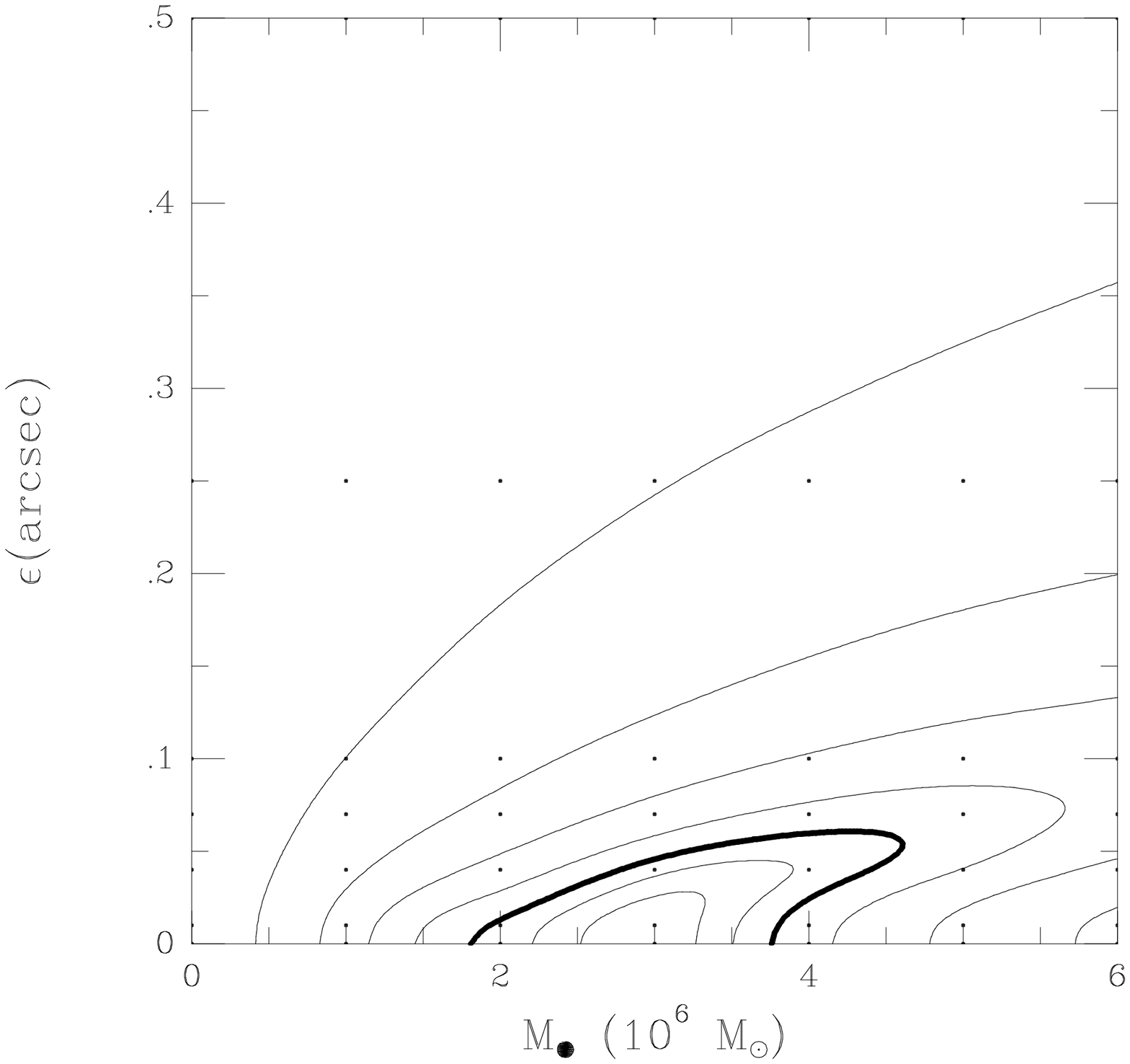}
\end{minipage}
\caption{\small\figcaptwo}
\end{figure}

%%% FIGURE 3 %%%

\begin{figure}
\epsfxsize=8.0truecm
\centerline{\epsfbox{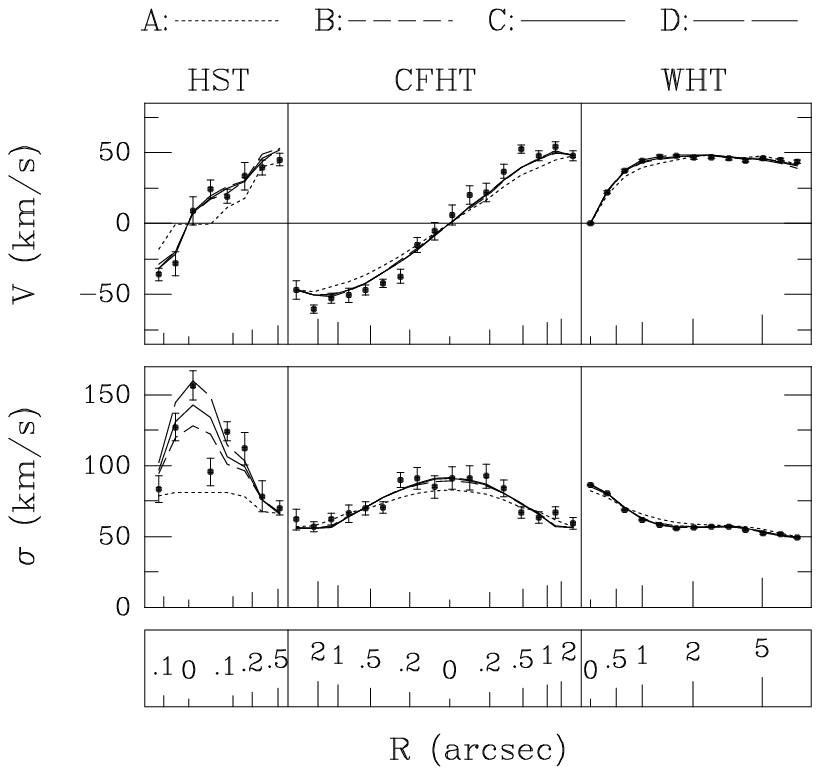}}
\caption{\small\figcapthree}
\end{figure}

\end{document}